# Modeling Iodine Deficiency

Paul Fong, Albert Li, Zoe Meadows, Rati Pillai, Chelsea Proutt, Alex Schneider

# Contents

## Executive Summary

This paper presents a four-unit, four-component mathematical model of iodine metabolism and its impact on thyroid hormone levels in the body. We focus on the relationships between iodine (I⁻), triiodothyronine (T3), thyroxine (T4), and thyroid-stimulating hormone (TSH) through the mixer, thyroid, sensor (pituitary gland), and metabolism. Iodine plays a fundamental role in maintaining metabolic homeostasis, as it is essential for the synthesis of T3 and T4, which regulate weight, energy, and other physiological functions. Iodine deficiency, which is one of the most common nutrient deficiencies in the world, can lead to hypothyroidism, a condition characterized by fatigue, weight gain, and cognitive impairments [1].

Our model tracks the movement of iodine through dietary intake, thyroid absorption, hormone synthesis, feedback regulation via TSH, and deiodization in metabolism. By evaluating different forms of the accounting equation governing these processes, we have concluded three key results. 1) Iodide availability directly impacts levels of T3 and T4 production, with both component flow rates declining at day 70 in a mildly diseased state, and day 60 in a severely diseased state. 2) TSH is an early diagnostic indicator of thyroid issues, with TSH levels rising to magnitudes of 25x in just ten days, nearly six times quicker than other biological indicators. 3) There is a 5 day difference in iodine storage depletion between mild and severe iodide deficiency. Understanding these results quantitatively provides insights into thyroid disorders and informs strategies for managing iodine deficiency on both individual and public health levels.

## Background

Iodine plays a critical role in maintaining normal metabolism in the human body through its involvement in thyroid hormone production. The thyroid gland uses iodine to synthesize two key hormones, triiodothyronine (T3) and thyroxine (T4) which work together to regulate weight, energy levels, internal temperature, and skin health, among other functions [2]. Since T3 and T4 are made from three and four iodine atoms, respectively, their production is dependent on iodine intake. T3 and T4 synthesis is regulated by thyroid-stimulating hormone (TSH), which increases when iodine intake is low and decreases when iodine intake is sufficient.

Iodine deficiency is caused by a diet with insufficient iodine intake and presents a significant public health concern. Iodine deficiency is more prevalent in mountainous areas and regions far from the ocean, where the soil is low in iodine, translating to a decrease in iodine concentration in the food sources [3]. Iodine deficiency is also a prevalent issue in developing countries without effective iodized salt programs [3]. Globally, approximately 30% of the population is at risk of iodine deficiency [1]. In the case of iodine deficiency, the thyroid gland is not able to produce enough T3 and T4, leading to hypothyroidism.

Hypothyroidism presents a significant health risk, as symptoms include fatigue, dry skin, weight gain, muscle weakness, irregular menstrual cycles, bradycardia, depression, and memory problems [5]. Adult hypothyroidism is associated with severe defects in intellect, fine motor skills, and balance, as well as deafness [2]. Hypothyroidism in infants is associated with poor growth, jaundice, umbilical hernia, and poor mental development. Over time, undiagnosed or untreated hypothyroidism can lead to high cholesterol and heart problems. The most common treatment for hypothyroidism is taking levothyroxine, which is a thyroid hormone medicine. Levothyroxine dosage is determined by TSH levels and must be precise, as overprescription of levothyroxine can cause sleep problems, shakiness, and in some cases heart palpitations. At the correct dosage, however, side effects are minimal [5].

Because iodine levels directly influence thyroid hormone production and TSH levels affect treatment dosage, it is crucial to quantify their interactions. The direct relationship between these molecules in the body makes it possible to model the impact of iodine levels over time. By focusing on the paths of iodine, T3, T4, and TSH in the body, a predictive model can be established to assess the progression of hypothyroidism and its metabolic consequences. This model allows for the simulation of varying iodine levels in the human body and displays the consequent effects of hypothyroidism in the long term.

## I.  Model

### A.  Biological Components

This model focuses on four biological components – Iodine (I⁻), triiodothyronine (T3), thyroxine (T4), and thyroid stimulating hormone (TSH). I⁻ is the fundamental element we are tracking in our model since our goal is to see the effects of iodine deficiency and its role in hypothyroidism. For our purposes, we assume iodide and iodine can be used interchangeably, as iodide is the form absorbed from the diet and is then converted to its active form iodine. We assume 100% of iodide is converted to iodine [29] (see Appendix A). We chose to focus on thyroid hormones T3 and T4 because they affect almost every cell in the body [6]. We also chose to track TSH in the body because it plays a crucial role in regulating T3 and T4 formation from I⁻. Tracking TSH levels is also important when it comes to prescribing the correct dosage of medicine.

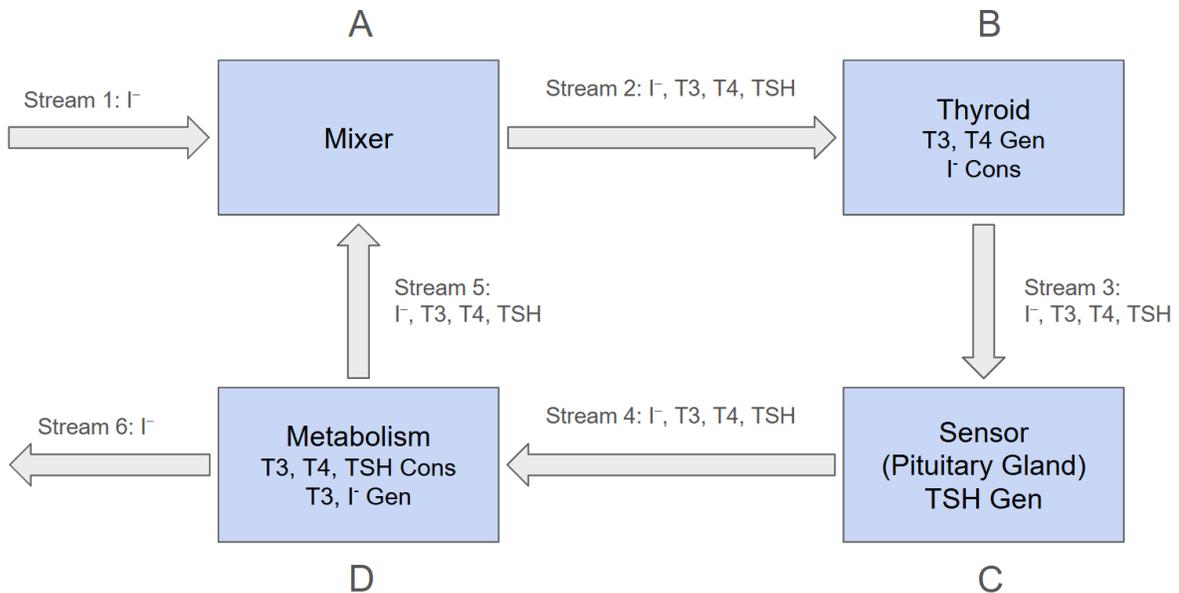

**Figure 1.** A four component model of the flow of iodine throughout the body.

To understand how iodine flows through the body, we developed a multicompartmental model to encompass all relevant areas where iodine is present within the body. Our model is shown in Figure 1, consisting of the four units essential to tracking iodine within the body. Iodine enters the model through stream 1 through dietary consumption, where it is mixed with recycled iodine, as well as other relevant compounds, such as T3, T4, and TSH.

These components are then sent to the thyroid via stream 2, where iodine is consumed to generate T3 and T4. The production of T3 and T4 is dependent on the concentration of TSH in the blood, so after flowing through the thyroid, the components flow to a sensor. The sensor represents the pituitary gland, which detects T3 and T4 levels, and regulates TSH production depending on these levels. T4 is then degraded into T3 and iodine via metabolism, while a small portion of these components is recycled back through the system via stream 5, where it joins the mixer described earlier. Nearly all of the intaken iodine within the body is excreted through urine, feces, sweat, or other methods.

Healthy, mild, and severe states of iodine deficiency are defined as 100%, 50% and 13% of an adequate 1190 nmol/day of iodine intake level. These values were derived from the diseased states of hypothyroidism as described by the World Health Organization [7].

### B.  Mixer (Unit A)

The "mixer" in this model represents the bloodstream. Iodine enters the bloodstream in Stream 1 through food such as seafood, dairy products, eggs, and iodized salt [4]. In our steady-state model, the flow rate of Stream 1 is dependent on the recommended iodine intake per day, as stated in literature [4]. This is the basis of our system. Under the dynamic disease states, the input is adjusted to reflect the relevant disease states, such as lowering the input of $I^-$ to reflect nutritional deficiencies.

$$\dot{n}_{1, I^-} = 1190 \text{ nmol/day} \qquad (1)$$

The second input stream to the mixer, Stream 5, consists of recycled $I^-$, T3, T4, and TSH that have already gone through the thyroid-pituitary gland-metabolism pathway and are ready for reuse. An explanation for the composition of this stream is provided in Section D (Metabolism).

The outlet, Stream 2, contains all these components and brings them from the mixer to the thyroid. The initial flow rate of $I^-$ in Stream 2 is determined by multiplying the rate of iodine consumption in the thyroid times five, as stated in literature [21]. After the initial values for all streams in the model were solved once, the subsequent flow rate of $I^-$ in Stream 2 was determined from the flow rate of $I^-$ entering the mixer (Streams 1 and 5). The flow rate of TSH in Stream 2 was calculated by multiplying the concentration of TSH in the blood with the blood flow rate entering the thyroid, with both values obtained from literature (see Appendix C). This method assumes there is a well-mixed concentration of TSH in the blood (see Appendix A). The flow rates of T3 and T4 were also obtained from a similar calculation; the typical concentration in blood was multiplied by the average blood flow rate to the thyroid (see Appendix C).

$$\dot{n}_{2,\,T3} = \dot{n}_{5,\,T3} \text{ nmol/day} \qquad (2)$$
$$\dot{n}_{2,\,T4} = \dot{n}_{5,\,T4} \text{ nmol/day} \qquad (3)$$
$$\dot{n}_{2,\,TSH} = \dot{n}_{5,\,TSH} \text{ mU/day} \qquad (4)$$
$$\dot{n}_{2,\,I^-} = \dot{n}_{5,\,I^-} + \dot{n}_{1,\,I^-} \text{ nmol/day} \qquad (5)$$

### C. Thyroid (Unit B)

Inlet Stream 2 provides the thyroid with all the components required to produce the appropriate amount of thyroid hormones for the body. Here, $I^-$ is consumed to generate hormones T3 and T4, based on the concentration of TSH. Additionally, storage of $I^-$ occurs in the thyroid [21]. The storage is not an additional unit, but an additional feature of the thyroid unit that allows for more accurate modeling of disease states. Under steady-state conditions, there is no change in the amount of $I^-$ in storage. However, under dynamic conditions, the amount of $I^-$ in storage changes as $I^-$ is released or stored, depending on the conditions in the body. Excess $I^-$ is stored until the maximum capacity of storage, as given by literature, is reached at 104157 nmol [18]. On the other hand, if TSH levels are above the steady-state TSH levels, then $I^-$ is secreted from storage, as detailed further on in this section (see Appendix C). All four components leave the thyroid, but the T3, T4, and $I^-$ are now in different proportions than when they entered via Stream 3. TSH remains constant. Therefore, the following governing equations are derived.

$$\dot{n}_{3,\,T3} = \dot{n}_{2,\,T3} + \dot{n}_{Gen,\,B,\,T3} \text{ nmol/day} \qquad (6)$$

$$\dot{n}_{3,\,T4} = \dot{n}_{2,\,T4} + \dot{n}_{Gen,\,B,\,T4} \text{ nmol/day} \qquad (7)$$

$$\dot{n}_{3,\,TSH} = \dot{n}_{2,\,TSH} \text{ mU/day} \qquad (8)$$

$$\dot{n}_{3,\,I^-} = \dot{n}_{2,\,I^-} + \dot{n}_{Cons,\,B,\,I^-} \text{ nmol/day} \qquad (9)$$

TSH stimulates the thyroid to secrete around 80% T4 and 20% T3 [6]. T3 contains three iodine molecules and is generally biologically active, while T4 contains four iodine molecules and is generally biologically inactive. The generation of T3 and T4 as a function of TSH is governed by several equations (see Appendix C for detailed derivations).

$$\dot{n}_{Gen,\,B,\,T4} = (0.502)\dot{n}_{2,\,TSH} \text{ nmol/day} \qquad (10)$$

$$\dot{n}_{Gen,\,B,\,T3} = (0.0324)\dot{n}_{2,\,TSH} \text{ nmol/day} \qquad (11)$$

$$\dot{n}_{Cons,\,B,\,I^-} = (3)\dot{n}_{Gen,\,B,\,T3} + (4)\dot{n}_{Gen,\,B,\,T4} \text{ nmol/day} \qquad (12)$$

Additional equations are used to calculate the I⁻ consumed to replenish the storage. If the storage is not full, then 10% of the I⁻ entering the thyroid is consumed and added to the storage.

$$\dot{n}_{Cons,\,B,\,I^-} = (3)\dot{n}_{Gen,\,B,\,T4} + (4)\dot{n}_{Gen,\,B,\,T3} + (0.1)\dot{n}_{2,\,I^-} \text{ nmol/day} \qquad (13)$$

Also, if 10% of the I⁻ entering the thyroid added to the current storage is greater than the maximum capacity, then only the necessary I⁻ to fill the storage is consumed. Finally, if the storage is already full, then no additional I⁻ is consumed (consumption fully governed by equation 12).

However, if the predicted consumption of I⁻ exceeds the body's maximum capacity, an alternative set of equations is used. This reflects the body's biological limits and I⁻ storage as determined from literature [21].

$$\dot{n}_{Cons,\,B,\,I^-} = (0.2)\dot{n}_{2,\,I^-} \text{ nmol/day} \qquad (14)$$

If there is no I⁻ in storage, then the following equations are used to calculate the generation of T3 and T4. The consumption of I⁻ is calculated using equation 14. Literature values for the ratio of T3 and T4 generation to TSH were normalized to the average flow rate in a health state to determine the appropriate numerical relationships.

$$\dot{n}_{Gen, B, T4} = (0.238)\dot{n}_{Cons, B, I^-} \text{ nmol/day} \qquad (15)$$

$$\dot{n}_{Gen, B, T3} = (0.0154)\dot{n}_{Cons, B, I^-} \text{ nmol/day} \qquad (16)$$

Otherwise, if there is I⁻ in storage, then the TSH levels entering the thyroid is compared with steady-state TSH levels to determine if I⁻ should be secreted from storage. If TSH levels entering the thyroid exceed the steady-state levels, then the generation of T3 and T4 is increased according to a log relationship determined from existing mathematical models in the literature [22][23][24][25]. The secretion follows a logarithmic relationship normalized to the average TSH in blood flow to the thyroid (see Appendix C).

$$\dot{n}_{Gen, B, T4} = (0.238) \{\dot{n}_{Cons, B, I^-} + (42.1)\log_{10}(\dot{n}_{2, TSH})\} \text{ nmol/day} \qquad (17)$$

$$\dot{n}_{Gen, B, T3} = (0.0154) \{\dot{n}_{Cons, B, I^-} + (42.1)\log_{10}(\dot{n}_{2, TSH})\} \text{nmol/day} \qquad (18)$$

Additionally, it is necessary to consider if the storage has sufficient capacity to secrete the required I⁻. If the I⁻ in storage is less than $(42.1)\log_{10}(\dot{n}_{2, TSH})$, which is the calculated secreted I⁻ from storage based on TSH levels, the actual I⁻ secreted from storage is simply all of the remaining I⁻ in storage.

### D. Sensor (Pituitary gland) (Unit C)

The sensor is the pituitary gland, which is located at the base of the brain and generates and releases TSH. The amount of TSH produced by the sensor is determined by the amount of T3 and T4 detected by the gland, where increased levels of T3 and T4 prevent the release of TSH via a negative feedback loop. The generation of TSH is calculated in relation to the presence of T4 in the preceding stream 3. When T3 and T4 levels drop, TSH is secreted to stimulate more production and maintain homeostasis. The steady-state relationship of T4 and TSH is used to extrapolate the dynamic generation (see Appendix C).

$$\dot{n}_{Gen, C, TSH} = (880953/\dot{n}_{3, T4}) \text{ mU/day} \qquad (19)$$

$$\dot{n}_{4, T3} = \dot{n}_{3, T3} \text{ nmol/day} \qquad (20)$$

$$\dot{n}_{4, T4} = \dot{n}_{3, T4} \text{ nmol/day} \qquad (21)$$

$$\dot{n}_{4, I^-} = \dot{n}_{3, I^-} \text{ nmol/day} \qquad (22)$$

$$\dot{n}_{4, TSH} = \dot{n}_{3, TSH} + \dot{n}_{Gen, C, TSH} \text{ mU/day} \qquad (23)$$

### E.   Metabolism (Unit D)

The metabolism represents the cells in the body that consume thyroid hormones to regulate their metabolic processes. Additionally, the degradation and processing of hormones in the kidney and liver are also represented in this unit. In the metabolism unit, the consumption of TSH is assumed to be the constant value solved in steady-state.

$$\dot{n}_{Cons, D, TSH} = 168 \text{ mU/day} \qquad (24)$$

Another important reaction is the deiodization of T4 to generate T3 and $I^-$. This reaction governs the generation of T3. Furthermore, a portion of T3 and T4 is fully consumed in the metabolism. This process also generates $I^-$. In summary, $I^-$ is generated in the metabolism by 3 major reactions: breaking down T3 completely (generates 3 $I^-$), breaking down T4 completely (generates 4 $I^-$), and breaking down T4 to T3 (generates 1 $I^-$). T4 is consumed to make T3, in addition to being completely broken down. The specific ratios were determined from steady-state relationships (see Appendix C).

$$\dot{n}_{Gen, D, T3} = (0.00720)\dot{n}_{4, T4} \text{ nmol/day} \qquad (25)$$

$$\dot{n}_{Cons, D, T4} = \dot{n}_{Gen, D, T3} + (0.0151)\dot{n}_{4, T4} \text{ nmol/day} \qquad (26)$$

$$\dot{n}_{Con, D, T3} = (0.139)\dot{n}_{4, T3} \text{ nmol/day} \qquad (27)$$

$$\dot{n}_{Gen, D, I^-} = (3)\dot{n}_{Con, D, T3} + (0.0605)\dot{n}_{4, T4} + \dot{n}_{Gen, D, T3} \text{ nmol/day} \qquad (28)$$

I⁻, T3, T4, and TSH are then sent through the bloodstream to be reused via Stream 5, and excess I⁻ is excreted via Stream 6. The values of stream 5 are governed by standard accounting equations that balance inputs, outputs, and the relevant generation or consumption terms. The only compound excreted from the body is I⁻. The specific ratio of I⁻ excreted was determined from steady-state relationships (see Appendix C).

$$\dot{n}_{5,\,TSH} = \dot{n}_{4,\,TSH} - \dot{n}_{Cons,\,D,\,TSH} \text{ mU/day} \tag{29}$$

$$\dot{n}_{5,\,T3} = \dot{n}_{4,\,T3} + \dot{n}_{Gen,\,D,\,T3} - \dot{n}_{Con,\,D,\,T3} \text{ nmol/day} \tag{30}$$

$$\dot{n}_{5,\,T4} = \dot{n}_{4,\,T4} - \dot{n}_{Cons,\,D,\,T4} \text{ nmol/day} \tag{31}$$

$$\dot{n}_{6,\,I^-} = (0.59)\,\dot{n}_{4,\,I^-} \text{ nmol/day} \tag{32}$$

$$\dot{n}_{5,\,I^-} = \dot{n}_{4,\,I^-} + \dot{n}_{Gen,\,D,\,I^-} - \dot{n}_{6,\,I^-} \text{ nmol/day} \tag{33}$$

# Results

## Key Result I: Iodine availability directly impacts levels of T3 and T4 production.

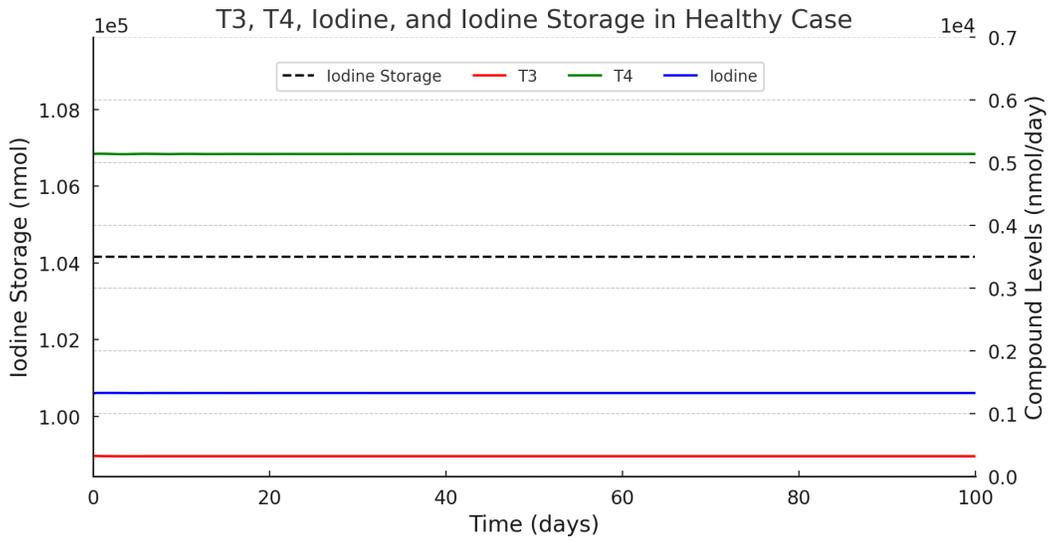

**Figure 2:** Iodine, T3, and T4 levels with healthy iodine intake in stream 5

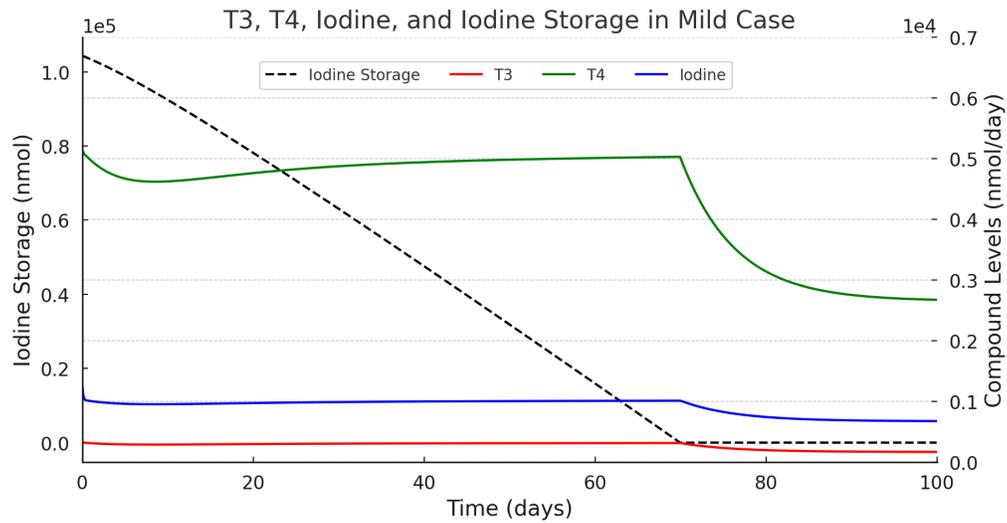

**Figure 3:** Iodine, T3, and T4 levels in a mild disease state in stream 5 (~50% normal intake levels)

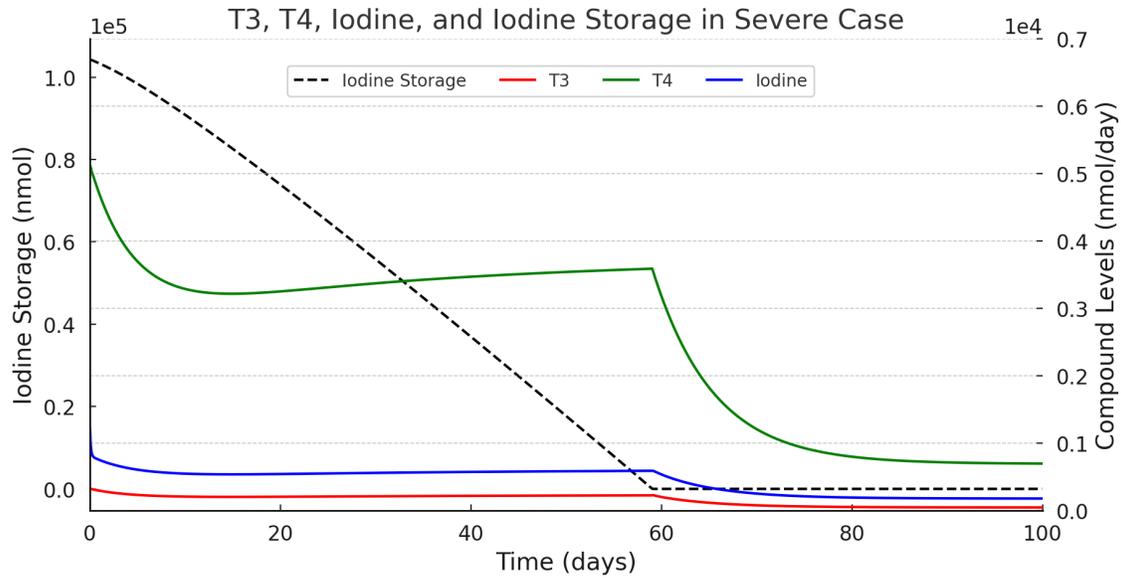

**Figure 4:** Iodine, T3, and T4 levels in a severe disease state in stream 5 (~13% normal intake levels)

Iodide uptake and storage in the thyroid is an essential process to allow for the conversion into T3 and T4 hormones [1]. Figure 2 displays the steady-state flow rates of T3 and T4 in stream 5, modeled with a healthy starting iodine intake level of 1190.47 nmol/day. Stream 5 represents blood composition after the metabolic generation and consumption of T3, T4 and TSH. The levels of T3 and T4 are shown to remain constant, indicating that there are sufficient levels of iodide in the system to maintain a consistent level of T3 and T4. In Figure 3 and Figure 4, which represent a mildly diseased state and severely diseased state respectively, it is clear that the insufficient levels of iodide in the system have decreased the flow rates of T3 and T4. In the mildly diseased state, the T3 and T4 drop off after 70 days, whereas in the severely diseased state, the decrease in the stream is sooner, after 60 days.

Prior to the drop in hormone flow rates, T3 and T4 in all three states are seen to maintain relatively steady stream levels from day 5 to 60. The drop offs in both disease state graphs also coincide with the depletion of iodine stores. This can be explained by the thyroid's attempt to compensate for the lack of iodide by increasing the secretion of iodide, which allows the thyroid to continue producing T3 and T4 for as long as possible [9]. Prolonged iodide deficiency then causes the rapid drop off of T4 hormone.  It has a similar effect on T3 hormone, which experiences a drop off starting at the same time, but declining much slower.  Both hormones drop off rates follow a logarithmic curve shape. These coinciding declines model the strong correlational relationships between iodide levels and T3 and T4 hormones. The body converts T4 into T3 in the body's cells [11], as modeled by the higher consumption term for T4 in the metabolism as compared to T3. This difference accounts for the more gradual decline of T3 levels in the diseased state [10]. However, it is clear that once T4 levels have fully depleted, so do T3 levels.

**Key Result II: TSH levels can be an early diagnostic indicator of thyroid issues.**

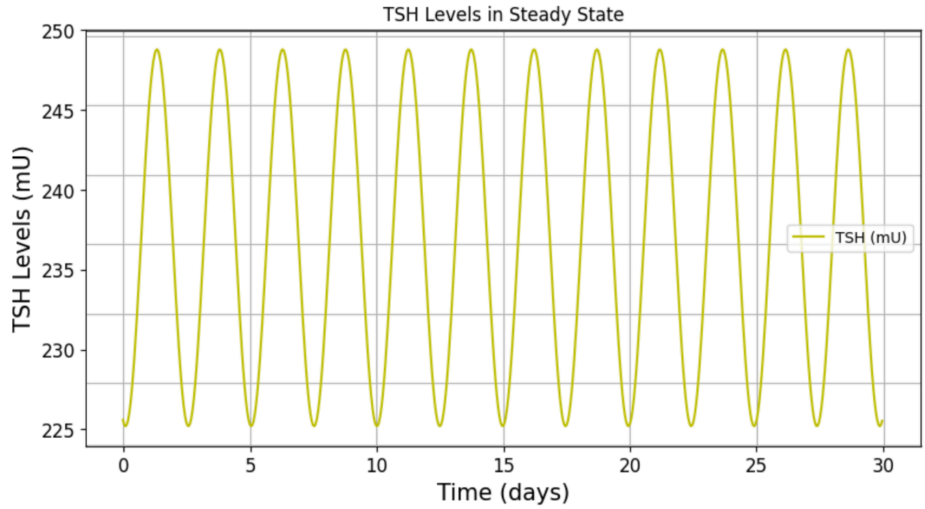

**Figure 5:** TSH Levels in steady-state in Stream 5

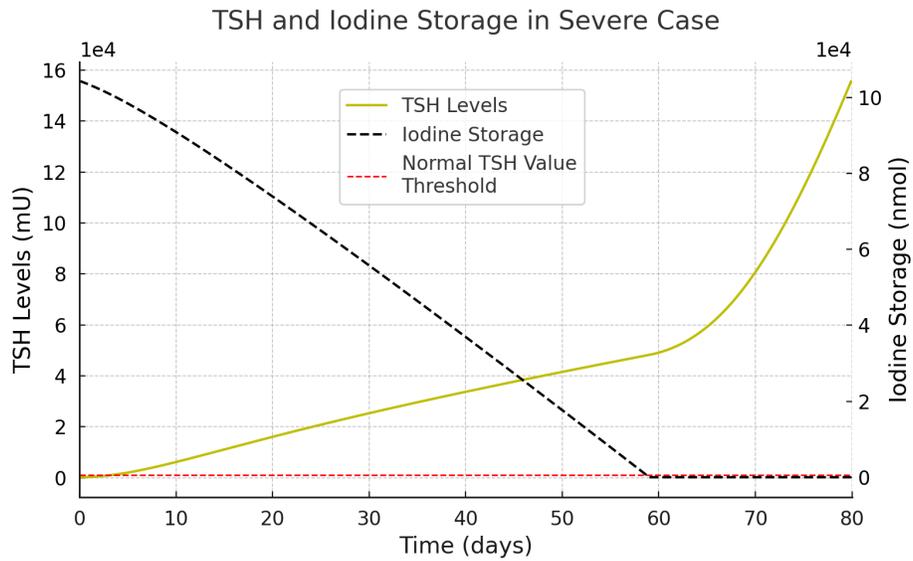

**Figure 6:** TSH Levels and Iodine Storage over time in Stream 5

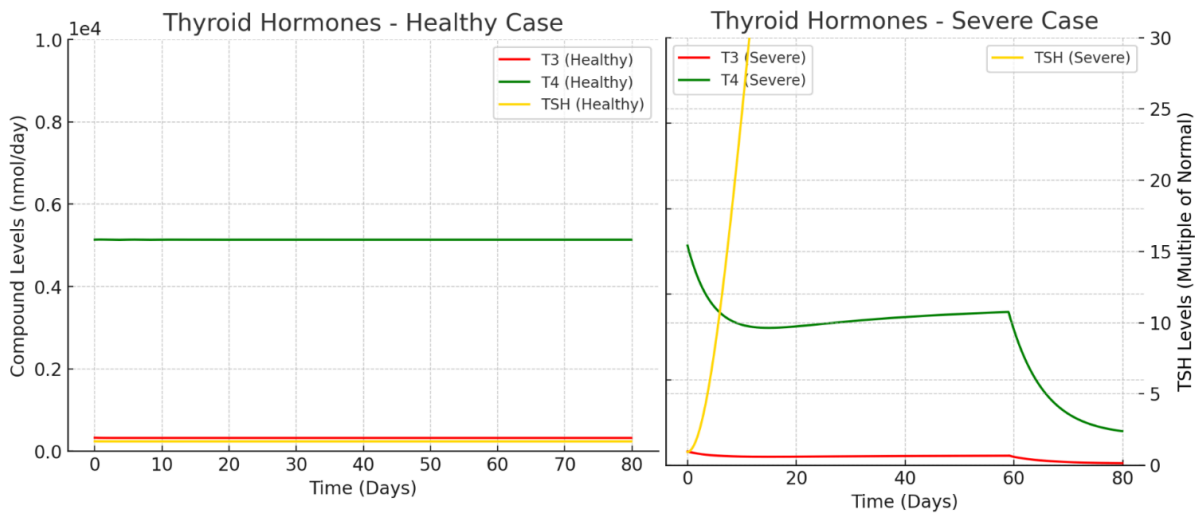

**Figure 7:** T3, T4, and TSH levels measured in Stream 5 in both a healthy and a severely diseased state.

Under steady-state conditions, TSH levels exhibit sinusoidal fluctuations in response to iodine intake, maintaining a range between approximately 230-250 mU (Figure 5). However, in a severely diseased state, as illustrated in Figure 6, TSH levels rapidly exceed this normal threshold. As iodine storage depletes over time, particularly after fully depleting in day 60, TSH levels surge dramatically in an effort to stimulate the thyroid gland to produce T3 and T4. Notably, this spike in TSH occurs well before any significant decline in other biochemical markers, such as T3 and T4 levels, highlighting its role as an early marker of iodine deficiency.

Figure 7 further supports this observation: TSH levels rise sharply from the onset of iodine depletion and exceed normal levels almost instantaneously, reaching magnitudes of 25x normal amounts by day 10. However, T3 and T4 levels do not sustain a noticeable decrease until around day 60, when iodine reserves are already nearly exhausted. This implies that discrepancies in TSH levels can be noticed 6 times faster compared to measuring for a significant decrease in T3 and T4 levels. This pattern also suggests that measuring TSH levels provides an earlier and more sensitive indication of iodine deficiency compared to direct T3 and T4 measurements. These findings can be backed from a biological standpoint as well. As the pituitary gland begins to receive slightly less T3 and T4 from the thyroid in the first 10 days, it will begin to produce an increasing amount of TSH on a noticeably significant level, which can be picked up quickly in diagnostic screenings.

From a clinical perspective, these findings emphasize the importance of early detection of thyroid dysfunction [14]. Monitoring TSH levels would allow for timely intervention, preventing the severe consequences of prolonged iodine deficiency and hypothyroidism. Routine TSH screenings, particularly in populations at risk for iodine deficiency, could facilitate early diagnosis and treatment, thereby reducing the prevalence and severity of thyroid-related complications.

**Key Result III: 10 Day Difference in Iodine Storage Depletion Between Mild and Severe Iodine Deficiency.**

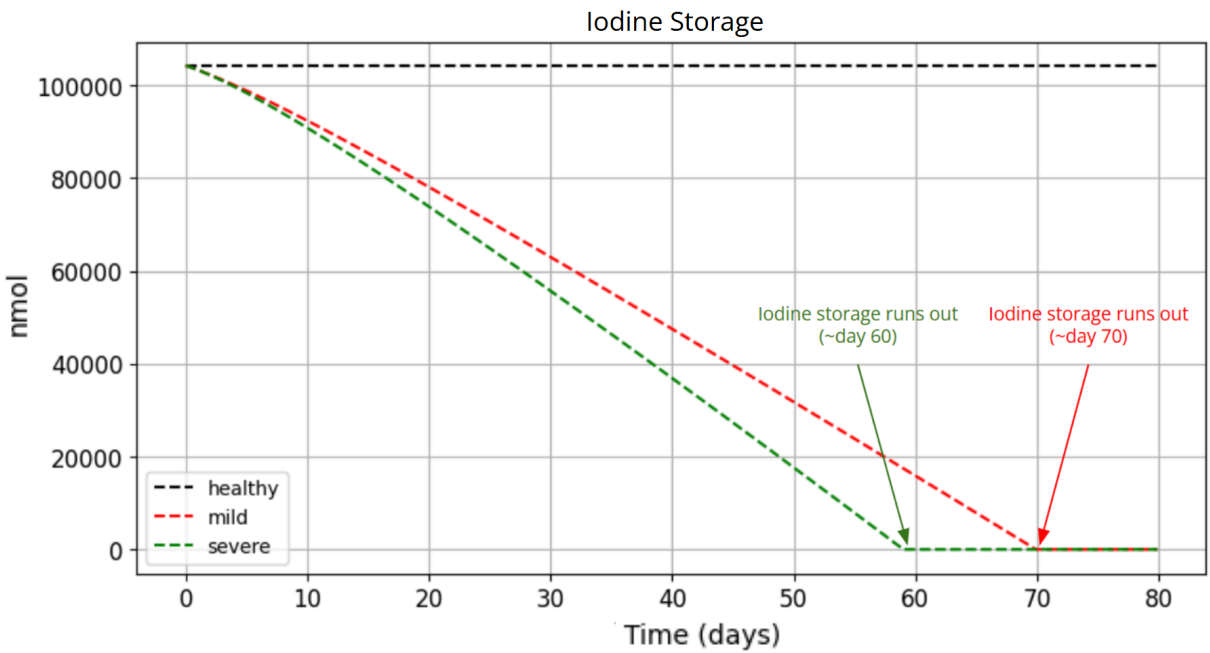

**Figure 8.** Storage of iodine in the thyroid over time for a healthy state, mildly diseased state, and severely diseased state.

Figure 8 shows that as iodine intake decreases, so does the time to deplete its storage in the thyroid. As iodine in the diet decreases, stream 1 of iodine decreases, which decreases the iodine in stream 2 going into the thyroid. The rest of the body needs the same amount of T3 and T4 to function normally, but the thyroid has less iodine to make these compounds. It uses iodine stores to compensate for the loss in intake. Storage levels drop quickly when iodine intake levels drop. When iodine intake is 595 nmol/day (50%), it takes 70 days in total for iodine stores to be completely used up. When iodine intake is at 155 nmol/day (13%) of the recommended daily value, it takes 60 days in total for iodine stores to be completely used up. This is a 10 day difference in iodine storage depletion between a mild and severe case of iodine deficiency. This depletion means that T3 and T4 cannot be made at all, and the metabolism cannot be regulated normally. TSH levels also rise rapidly, indicating hypothyroidism as discussed previously in the background. Decreasing iodine intake quickly affects the thyroid and its ability to make critical hormones that maintain metabolic rates in the rest of the body.

## Discussion

### A. Impacts of Iodine Deficiency on Hormone Production

Iodine is essential for thyroid hormone production and the regulation of metabolic processes. As modeled in our steady-state system (see Figure 2), sufficient iodine intake facilitates a consistent production of T3 and T4 hormones, which play key roles in regulating weight, energy levels, internal temperature, and more [3]. In contrast, in an iodine-deficient state, the body is unable to produce T3 and T4 hormones at a regulated state. In a mildly diseased model, where iodine intake is 50% of an adequate iodine intake, it is clear that the thyroid uses stored iodine levels to temporarily maintain hormone production at around $0.5 \times 10^5$ nmol/day of T4 and $0.3 \times 10^5$ nmol/day of T3 (see Fig 3) before rapid depletion of T4 is observed at 70 days (see Figure 3). T3 declines at a slower rate due to the T4 to T3 conversions that occur in the body's cells (represented by the metabolism unit). In the severely diseased model, where iodine intake is 13% of an adequate intake level, the same phenomena was observed, but at an accelerated rate. T4 drops from around $0.35 \times 10^5$ nmol/day to $0.8 \times 10^5$ nmol/day from day 60 to day 100, while T3 drops from around $0.25 \times 10^5$ nmol/day to almost 0 nmol/day from day 60 to day 100.

In response to the depletion of T3 and T4 in the severe disease state, TSH levels are shown to increase at around 830 mU/day until day 60, then at a higher rate from day 60 to 80 (see Figure 6). This increase in rate coincides with the drainage of iodine stores in the thyroid as the pituitary gland senses the deficiency and increases TSH production. In increasing levels of iodine deficiency, the exhaustion rate of thyroid iodine stores is proportionally increasing. This is an important result as elevated TSH acts as an early diagnostic marker for iodine deficiency and hypothyroidism. In severe cases of hypothyroidism, where T3 and T4 hormones reach extremely low levels, catastrophic metabolic and cardiovascular effects can result in death.

**B. Strengths and Weaknesses of Model**

One strength of our model is the ability to track the movement of iodine, T3, T4, and TSH through the thyroid, pituitary gland, metabolism, and rest of the body. We were able to derive equations that relate the flow of all of these molecules to one another in order to get reasonable quantitative and qualitative results. The model and the equations we used are intuitive to understand and balance simplicity with complexity. For example, we were able to include all of the important organs and compounds that affect the system we were modeling without having extraneous units due to the natural complexity of biological systems such as this one.

Another strength of our model is that our steady-state calculations are qualitatively and quantitatively accurate. For example, the stream 2 iodine mole flow rate can be checked using both a mass balance of the streams, and a relationship found in literature (see discussion of stream 2 in model analysis) where $\dot{n}_{2,\text{I-}} = \dot{n}_{5,\text{I-}} + \dot{n}_{1,\text{I-}}$ nmol/day (Eq 5). When comparing these two values, there is a minimal error of 0.38%, which is acceptable given the assumptions made to accommodate for biological factors and variation between individuals in the population. The flow rates of iodine, T3, T4, and TSH through the body also match literature values, and our calculations balanced out correctly. Our model predicts that T3 and T4 flow rates begin to deplete after 60-70 days of iodine deficiency, coinciding with the depletion of iodine storage. This aligns with literature values as the human thyroid is known to store approximately 15-20 mg of iodine and have a daily iodine turnover of 60-95 μg [14]. These reserves could sustain a normal thyroid function for around 2-3 months, aligning with our model results of 60-70 days.

In addition, our model allows for any adjustment in iodine input to be tracked. This helps demonstrate different levels of iodine deficiency and its effects on the flow of iodine, T3, T4, and TSH through the body along with the amount of iodine being stored in the thyroid. This model supports an important global health concern that affects 2 billion people and can be used to help understand daily recommended iodine values better and more accurately.

However, our model has many restrictions due to assumptions and the nature of our approach. One of these limitations is that there was a lack of accurate, high-quality measurements of the compounds we were tracking in the human body. Discrepancies in literature data limit our ability to analyze the precision of our model to the highest extent, especially as it cannot account for individual physiological variations. However, we evaluated the literature values we found in order to choose the most reasonable and consistent values to use in our calculations. We assumed we were modeling a middle aged man, which limits the applicability of this model to other groups such as women and children. It should also be noted that there are a lot of fluctuating values in the compounds we were tracking throughout time and across different people, even ones of similar demographics.

We had to make many assumptions about our dynamic system in order to reduce its complexity. We assumed linear relationships and direct proportions between the generation of T3 and T4 based on TSH levels in the thyroid along with the generation of TSH based on T4 levels in the pituitary gland. Real body systems are more complex with compounding variables so this is not the most accurate representation of the actual biological system.

Furthermore, we assumed that T3, T4, and TSH consumption in the metabolism were constant values that our steady-state model calculated for us. In a real body, the metabolism would likely attempt to maintain homeostasis by slowing down and consuming less of these hormones as iodine stores decreased. However, this would be a very complex system to model, and the lack of literature data on this relationship led us to make this assumption.

Another limitation of our dynamic system is that we used Euler's Method in our calculations. This algorithm is highly dependent on the initial values (which we calculated from our steady-state model) and the step size. This often leads to fluctuations in results, which could explain some of the behavior of the results the model produced. Euler's Method is also less accurate at times far from the initial value. Long term extrapolation of our data is not a good representation of the system we modeled because it continues to increase or decrease depending on the specific compound concentrations, while in an actual human body, other compounds and systems get involved to maintain homeostasis.

## Conclusion

This model, while facing some limitations, effectively predicts the impacts of iodine intake on the production of T3 and T4 hormones. The model incorporates 4 units which account for biological processes and includes the key organs responsible for processing iodine.

Our results highlight that iodine deficiency disrupts the body's ability to effectively produce and maintain T3 and T4 hormones, resulting in a diseased state of hypothyroidism. Although the thyroid is able to compensate temporarily using existing iodine storage, long term it is unable to prevent a decline in thyroid hormone production. The accelerated exhaustion of iodine stores in more severe deficiency models highlight's the body's limited capacity to adapt to sustained iodine scarcity.

Although our model utilizes assumptions and is not an accurate representation for each individual's variable metabolic rates, it holds value as it is able to model the relationships between iodine intake, T3 and T4 hormone production, and TSH regulation. By capturing the dynamic interactions between these components, the model gives insight into the effects of prolonged iodine deficiency, as well as displaying the relative rates of depletion based on varying levels of iodine deficiency. The observed rise in TSH serves as an indicator of hypothyroidism and could be used for early diagnosis in patients.

# Appendix A: Assumptions

1. All iodine from food is absorbed by the body.
2. 100% iodide absorbed from the diet is converted to its active form iodine. Iodide and iodine can be used interchangeably for the purposes of this paper. [29]
3. $I^-$, T3, T4, and TSH are the only relevant components in the iodine pathway system. [6]
4. The only relevant places of generation, consumption, and mixing of the four listed components are as follows:
   a. Mixer – The mixer is the bloodstream. $I^-$ enters from food, all four components enter via metabolism, and all four components leave.
   b. Thyroid – All four components enter and leave. Only T3 and T4 are generated, while only $I^-$ is consumed.
   c. Sensor (Pituitary Gland) – All four components enter and leave. Only TSH is generated.
   d. Metabolism – All components that don't leave the body are reused by the mixer.
5. There is no accumulation of components in a healthy, steady-state individual.
6. There is a well-mixed concentration of components in the bloodstream.

# Appendix B: Additional Tables

Table B1: Descriptions of all streams in model.

| Description | Symbol | Value | Source |
|---|---|---|---|
| Iodine entering body | $\dot{n}_{1,\,I^-}$ | 1190.47 nmol/day | [4] |
| Iodine entering thyroid | $\dot{n}_{2,\,I^-}$ | 2496.14 nmol/day | [21] (see Appendix C calculations) |
| T3 entering thyroid | $\dot{n}_{2,\,T3}$ | 323.07 nmol/day | [16] and [17] (see Appendix C |

| | | | calculations) |
|---|---|---|---|
| T4 entering thyroid | $\dot{n}_{2, T4}$ | 5133.6 nmol/day | [16] and [17] (see Appendix C calculations) |
| TSH entering thyroid | $\dot{n}_{2, TSH}$ | 237 nmol/day | [16] and [24] (see Appendix C calculations) |
| T3 generated by thyroid | $\dot{n}_{Gen, B, T3}$ | 7.6809 nmol/day | [12] |
| T4 generated by thyroid | $\dot{n}_{Gen, B, T4}$ | 119.047 nmol/day | [13] |
| Iodine consumed in thyroid | $\dot{n}_{Cons, B, I-}$ | 499.228 nmol/day | [19][20] (solved steady-state mass balance using known stoichiometric ratios) |
| Iodine leaving thyroid | $\dot{n}_{3, I-}$ | 1996.91 nmol/day | Solved steady-state mass balance |
| T3 leaving thyroid | $\dot{n}_{3, T3}$ | 330.75 nmol/day | Solved steady-state mass balance |
| T4 leaving thyroid | $\dot{n}_{3, T4}$ | 5332.647 nmol/day | Solved steady-state mass balance |
| TSH leaving thyroid | $\dot{n}_{3, TSH}$ | 237 mU/day | Solved steady-state mass balance |
| TSH generated by pituitary gland | $\dot{n}_{Gen, C, TSH}$ | 165.2 mU/day | [26] See Appendix C for detailed analysis and justification with regards to TSH consumption. |
| Iodine leaving pituitary gland | $\dot{n}_{4, I^-}$ | 1996.91 nmol/day | Solved steady-state mass balance |
| T3 leaving pituitary gland | $\dot{n}_{4, T3}$ | 330.75 nmol/day | Solved steady-state mass balance |
| T4 leaving pituitary gland | $\dot{n}_{4, T4}$ | 5332.647 nmol/day | Solved steady-state mass balance |
| TSH leaving pituitary gland | $\dot{n}_{4, TSH}$ | 402.2 mU/day | Solved steady-state mass balance |
| Iodine produced from T3 deiodization, T4 deiodization, and complete breakdown of T3 and T4 | $\dot{n}_{Gen, D, I^-}$ | 499.228 nmol/day | Solved steady-state mass balance |
| T3 produced in metabolism from T4 deiodization | $\dot{n}_{Gen, D, T3}$ | 38.402 nmol/day | [28] |
| T3 consumption in metabolism | $\dot{n}_{Cons, D, T3}$ | 46.0829 nmol/day | [27] |

| T4 deiodization and complete consumption via metabolism | $\dot{n}_{Cons, D, T4}$ | 119.047 nmol/day | Solved steady-state mass balance |
|---|---|---|---|
| TSH consumed in metabolism | $\dot{n}_{Cons, D, TSH}$ | 168 nmol/day | Solved steady-state mass balance. See Appendix C for detailed analysis and justification with regards to TSH generation. |
| Iodine reused after metabolism | $\dot{n}_{5, I^-}$ | 1305.66 nmol/day | Solved steady-state mass balance |
| T3 reused after metabolism | $\dot{n}_{5, T3}$ | 323.07 nmol/day | [16] and [17] (see Appendix C calculations) |
| T4 reused after metabolism | $\dot{n}_{5, T4}$ | 5133.6 nmol/day | [16] and [17] (see Appendix C calculations) |
| TSH reused after metabolism | $\dot{n}_{5, TSH}$ | 237 mU/day | [16] and [24] (see Appendix C calculations) |
| Iodine leaving body | $\dot{n}_{6, I^-}$ | 1190.47 nmol/day | Solved steady-state mass balance |

Table B2: Reference Values

| Compound | Value | Source |
|---|---|---|
| T3 | MW: 651 g/mol | [19] |
| T4 | MW: 777 g/mol | [20] |
| I- | MW: 126.9045 g/mol | [8] |
| TSH | MW: 28,000 g/mol | [15] |
| Average blood flow to thyroid | 31 mL/min | [16] |
| T3 concentration in blood | 80-220 ng/dL | [17] |
| T4 concentration in blood | 5-12 µg/dL | [17] |
| Thyroid storage capacity of I- | 104,167 nmol | [18] |
| TSH concentration in blood | 5.3 µU/mL | [24] |

## Appendix C: Additional Equations and Calculations

Steady-state Calculations:
- The steady-state model was initially solved by hand to determine if the existing literature values were sufficient to completely and accurately solve the necessary equations. To verify the results, Python was used (code in Appendix E). The steady-state model was run over 100 days, and the final values are summarized in the tables above (Appendix B). It is important to note some minor discrepancies between the initial accounting equations and the final, solved values. This apparent discrepancy can be explained by the assumptions and literature data required to develop this model. Since averages of many data points are used, it is expected that the calculations will not perfectly line up.

Dynamic Model Calculations:
- The relevant calculations for each stream and unit are described in this section. Detailed calculations are included for any coefficients or constants that need to be derived. Equations pertaining to the standard accounting equations are not restated (see the relevant discussion in the Model section).

Stream 2 Calculations:

To determine the flow rates of TSH, T3, and T4 to the thyroid ($\dot{n}_{2, T3}$, $\dot{n}_{2, T4}$, $\dot{n}_{2, TSH}$). The concentration of each compound in the blood was multiplied by the average blood flow to the thyroid. It was assumed that the concentration of all three compounds in the body are well-mixed, so the typical concentration is considered representative of the flow rate to the thyroid specifically.

- Average blood flow to thyroid = 31 mL/min = 44640 mL/day [16]
- T3:
  - Use average concentration in blood: 80-220 ng/dL => 150 ng/dL [17]
  - Multiply by average blood flow to thyroid: 150 ng/dL * 44640 mL/day = 6.696 e-4 g/day = 1020.7 nmol/day
- T4:
  - Use average concentration in blood: 5-12 µg/dL => 8.5 µg/dL [17]
  - Multiply by average blood flow to thyroid: 8.5 µg/dL * 44640 mL/day = 3.79 e-3 g/day = 48833 nmol/day

Stream 3 Calculations:
- Equation 10: assume the steady state ratio between $\dot{n}_{Gen, B, T4}$ and $\dot{n}_{2, TSH}$ governs relationship in dynamic conditions
  - $\dot{n}_{Gen, B, T4}$ = 119.047 nmol/day
  - $\dot{n}_{2, TSH}$ = 237 mU/day

- Ratio of $\dot{n}_{Gen,\,B,\,T4}$ / $\dot{n}_{2,\,TSH}$ = 0.502 nmol/mU

$$\dot{n}_{Gen,\,B,\,T4} = (0.502)\dot{n}_{2,\,TSH} \text{ nmol/day} \quad (10)$$

- Equation 11: assume the steady state ratio between $\dot{n}_{Gen,\,B,\,T3}$ and $\dot{n}_{2,\,TSH}$ governs relationship in dynamic conditions
  - $\dot{n}_{Gen,\,B,\,T3}$ = 7.6809 nmol/day
  - $\dot{n}_{2,\,TSH}$ = 237 mU/day
  - Ratio of $\dot{n}_{Gen,\,B,\,T3}$ / $\dot{n}_{2,\,TSH}$ = 0.0324 nmol/mU

$$\dot{n}_{Gen,\,B,\,T3} = (0.0324)\dot{n}_{2,\,TSH} \text{ nmol/day} \quad (11)$$

- Equation 12: the consumption of I⁻ is governed by the stoichiometric requirements of T3 and T4. T3 requires 3 I⁻ and T4 requires 4 I⁻ [19][20]

$$\dot{n}_{Cons,\,B,\,I\text{-}} = (3)\dot{n}_{Gen,\,B,\,T4} + (4)\dot{n}_{Gen,\,B,\,T3} \text{ nmol/day} \quad (12)$$

- Equation 15: assume the steady state ratio between $\dot{n}_{Gen,\,B,\,T4}$ and $\dot{n}_{Cons,\,B,\,I\text{-}}$ governs relationship in dynamic conditions
  - $\dot{n}_{Gen,\,B,\,T4}$ = 119.047 nmol/day
  - $\dot{n}_{Cons,\,B,\,I\text{-}}$ = 499.228 nmol/day
  - Ratio of $\dot{n}_{Gen,\,B,\,T4}$ / $\dot{n}_{Cons,\,B,\,I\text{-}}$ = 0.238

$$\dot{n}_{Gen,\,B,\,T4} = (0.238)\dot{n}_{Cons,\,B,\,I\text{-}} \text{ nmol/day} \quad (15)$$

- Equation 16: assume the steady state ratio between $\dot{n}_{Gen,\,B,\,T3}$ and $\dot{n}_{Cons,\,B,\,I\text{-}}$ governs relationship in dynamic conditions
  - $\dot{n}_{Gen,\,B,\,T3}$ = 7.6809 nmol/day
  - $\dot{n}_{Cons,\,B,\,I\text{-}}$ = 499.228 nmol/day
  - Ratio of $\dot{n}_{Gen,\,B,\,T4}$ / $\dot{n}_{Cons,\,B,\,I\text{-}}$ = 0.0154

$$\dot{n}_{Gen,\,B,\,T3} = (0.0154)\dot{n}_{Cons,\,B,\,I\text{-}} \text{ nmol/day} \quad (16)$$

- Equation 17 and Equation 18:
  - The secretion of I⁻ from storage is governed by a log-linear relationship determined by the underlying kinetic and enzymatic principles. Due to the additional complexity involved, the math was appropriately simplified to match the scope of this model. The secretion can be accurately modeled as $\log_{10}(\dot{n}_{2,\,TSH})$ [22][23][24][25].

- ○ However, the TSH levels have to be normalized to the average, steady-state TSH value. This is done by introducing a coefficient: $42.1 = 100/\log(237)$, where 237 nmol/day is the steady-state TSH flow rate out of the thyroid.
- ○ The secreted I⁻ was introduced mathematically to the $\dot{n}_{Cons,\,B,\,I^-}$ term in equations 15 and 16, respectively for equations 17 and 18. Therefore, the secretion can be accurately modeled as $(42.1)\log_{10}(\dot{n}_{2,\,TSH})$.

$$\dot{n}_{Gen,\,B,\,T4} = (0.238)\,\{\dot{n}_{Cons,\,B,\,I^-} + (42.1)\log_{10}(\dot{n}_{2,\,TSH})\}\ \text{nmol/day} \qquad (17)$$

$$\dot{n}_{Gen,\,B,\,T3} = (0.0154)\,\{\dot{n}_{Cons,\,B,\,I^-} + (42.1)\log_{10}(\dot{n}_{2,\,TSH})\}\text{nmol/day} \qquad (18)$$

Stream 4 Calculations:

- The generation of TSH in the pituitary gland was determined via the steady-state model. Under steady state conditions:
  - ○ $\dot{n}_{Gen,\,C,\,TSH} = 165.2$ mU/day
  - ○ $\dot{n}_{3,\,T4} = 5332.647$ nmol/day
  - ○ Product of $\dot{n}_{Gen,\,C,\,TSH} * \dot{n}_{3,\,T4} = 880953$ nmol*mU/day$^2$
- Extrapolate the steady-state relationship to determine the overall equation:

$$\dot{n}_{Gen,\,C,\,TSH} = (880953/\dot{n}_{3,\,T4})\ \text{mU/day} \qquad (19)$$

Stream 5 Calculations:

- Equation 24: the consumption of TSH is initially calculated using a steady-state mass balance accounting equation.
  - ○ $\dot{n}_{Gen,\,C,\,TSH} = 165.2$ mU/day [26]
  - ○ Thus, the initial $\dot{n}_{Cons,\,D,\,TSH} = 165.2$ mU/day so that the accumulation of TSH in the body is zero. However, as the steady state model is solved over time, the consumption of TSH must approach 168 mU/day to maintain zero accumulation. Therefore, the ultimate steady state consumption is 168 mU/day. This apparent discrepancy can be explained by the assumptions and literature data required to develop this model. Since averages of many data points are used, it is expected that the calculations will not perfectly line up.

$$\dot{n}_{Cons, D, TSH} = 168 \text{ mU/day} \qquad (24)$$

- Equation 25: assume the steady state ratio between $\dot{n}_{Gen, D, T3}$ and $\dot{n}_{4, T4}$ governs relationship in dynamic conditions
  - ○ $\dot{n}_{Gen, D, T3} = 38.402 \text{ nmol/day}$
  - ○ $\dot{n}_{4, T4} = 5332.647 \text{ nmol/day}$
  - ○ Ratio of $\dot{n}_{Gen, D, T3} / \dot{n}_{4, T4} = 0.00720$

$$\dot{n}_{Gen, D, T3} = (0.00720)\dot{n}_{4, T4} \text{ nmol/day} \qquad (25)$$

- Equation 26: assume the steady state ratio between $\dot{n}_{Cons, D, T4}$, $\dot{n}_{Gen, D, T3}$, and $\dot{n}_{4, T4}$ governs relationship in dynamic conditions
  - ○ $\dot{n}_{Cons, D, T4} = 119.047 \text{ nmol/day}$
  - ○ $\dot{n}_{Gen, D, T3} = 38.402 \text{ nmol/day}$
  - ○ $\dot{n}_{4, T4} = 5332.647 \text{ nmol/day}$
  - ○ Ratio of $(\dot{n}_{Cons, D, T4} - \dot{n}_{Gen, D, T3}) / \dot{n}_{4, T4} = 0.0151$

$$\dot{n}_{Cons, D, T4} = \dot{n}_{Gen, D, T3} + (0.0151)\dot{n}_{4, T4} \text{ nmol/day} \qquad (26)$$

- Equation 27: assume the steady state ratio between $\dot{n}_{Con, D, T3}$ and $\dot{n}_{4, T3}$ governs relationship in dynamic conditions
  - ○ $\dot{n}_{Con, D, T3} = 46.0829 \text{ nmol/day}$
  - ○ $\dot{n}_{4, T3} = 330.75 \text{ nmol/day}$
  - ○ Ratio of $\dot{n}_{Con, D, T3} / \dot{n}_{4, T3} = 0.139$

$$\dot{n}_{Con, D, T3} = (0.139)\dot{n}_{4, T3} \text{ nmol/day} \qquad (27)$$

- Equation 28: assume the steady state ratio between $\dot{n}_{Gen, D, I^-}$, $\dot{n}_{Con, D, T3}$, $\dot{n}_{4, T4}$, and $\dot{n}_{Gen, D, T3}$ governs the relationship in dynamic conditions
    - The consumption of T3 generates 3 $I^-$, so the $\dot{n}_{Con, D, T3}$ terms is multiplied by 3.
    - T4 entering the metabolism is consumed in two ways: generation of T3 and complete degradation. This relationship is fully explained in equation 26. Therefore, since $(0.0151)\dot{n}_{4, T4}$ represents the T4 being completely degraded, it also represents the $I^-$ generated from T4 degradation. Since there are 4 $I^-$ generated from each T4 broken down, the $\dot{n}_{4, T4}$ term contributes $4*0.0151 = 0.0605$.
    - Finally, the generation of T3 also directly corresponds with the generation of $I^-$, since the T3 is generated from T4, which yields one $I^-$. Therefore, the $\dot{n}_{Gen, D, T3}$ directly contributes to $\dot{n}_{Gen, D, I^-}$.

$$\dot{n}_{Gen, D, I^-} = (3)\dot{n}_{Con, D, T3} + (0.0605)\dot{n}_{4, T4} + \dot{n}_{Gen, D, T3} \text{ nmol/day} \qquad (28)$$

- Equation 32: assume the steady state ratio between $\dot{n}_{6, I^-}$ and $\dot{n}_{4, I^-}$ governs relationship in dynamic conditions
    - $\dot{n}_{6, I^-} = 1190.47$ nmol/day
    - $\dot{n}_{4, I^-} = 1996.91$ nmol/day
    - Ratio of $\dot{n}_{6, I^-}/ \dot{n}_{4, I^-} = 0.59$

$$\dot{n}_{6, I^-} = (0.59) \, \dot{n}_{4, I^-} \text{ nmol/day} \qquad (32)$$

## Appendix D: Documentation of Artificial Intelligence Use

Throughout the course of this project, our team utilized the artificial intelligence tools at our disposal. In particular, ChatGPT Model 4o was used in multiple steps of the process. For example, when initially brainstorming which compound to build our project around, we compiled a list of 40 or so possible nutrients, then asked ChatGPT to give us a few effects of both too much and too little of the nutrient. Since it would have taken a lot of time to perform thorough research on this many nutrients, this allowed us to get a brief overview of many nutrients, from which we were able to incorporate this information and filter our list to 5-8 nutrients that we could do preliminary research on, and finally decide on our final choice of nutrient. In addition, while coding our dynamic model, we ran into a multitude of errors, as well as a few graph formatting errors, so by using ChatGPT's debugging abilities, we were able to smooth out our code and generate consistent, relevant plots.

**Appendix E: Code**

```python
#import libraries
import matplotlib.pyplot as plt
import matplotlib as mpl
import numpy as np
#from scipy.integrate import odeint

#make time array
t0=0
tf = 80
dt=0.1
t = np.arange(t0, tf, dt)
#make iodine input array
fraction = 0.50
I = fraction * 1200 * np.ones(len(t))

#define normal conditions
SSTSH = 237 ##steady state TSH

#make empty arrays to store values in and set initial values
#iodine
n_IN = np.zeros(len(t))
n1_I = np.zeros(len(t))
n2_I = np.zeros(len(t))
conB_I = np.zeros(len(t))
```

```python
n3_I = np.zeros(len(t))
n4_I = np.zeros(len(t))
genD_I = np.zeros(len(t))
n5_I = np.zeros(len(t))
n6_I = np.zeros(len(t))
nstorage_I = np.zeros(len(t))

n_IN[0] = 2496
n1_I[0] = I[0]  #nmol/day
n2_I[0] = 2496 #nmol/day
conB_I[0] = 499.228  #nmol/day
n3_I[0] = n2_I[0] - conB_I[0] #nmol/day
n4_I[0] = n3_I[0]  #nm/day
genD_I[0] = 499.228 #nmol/day
n6_I[0] = 1190.47 #nmol/day
n5_I[0] = n4_I[0] - n6_I[0] + genD_I[0] #nmol/day
nstorage_I[0] = 104167 #nmol

#T4
n2_T4 = np.zeros(len(t))
genB_T4 = np.zeros(len(t))
n3_T4 = np.zeros(len(t))
n4_T4 = np.zeros(len(t))
conD_T4 = np.zeros(len(t))
n5_T4 = np.zeros(len(t))

n2_T4[0] = 5136.6 #nmol/day
genB_T4[0] = 119.047 #nmol/day
n3_T4[0] = n2_T4[0] + genB_T4[0] #nmol/day
n4_T4[0] = n3_T4[0] #nmol/day
conD_T4[0] = 119.047 #nmol/day
```

```python
n5_T4[0] = n4_T4[0] - conD_T4[0] #nmol/day

#T3
n2_T3 = np.zeros(len(t))
genB_T3 = np.zeros(len(t))
n3_T3 = np.zeros(len(t))
n4_T3 = np.zeros(len(t))
genD_T3 = np.zeros(len(t))
conD_T3 = np.zeros(len(t))
n5_T3 = np.zeros(len(t))

n2_T3[0] = 323.07 #nmol/day
genB_T3[0] = 7.68 #nmol/day
n3_T3[0] = n2_T3[0] + genB_T3[0] #nmol/day
n4_T3[0] = n3_T3[0] #nmol/day
genD_T3[0] = 38.402 #nmol/day
conD_T3[0] = 46.083 #nmol/day
n5_T3[0] = n4_T3[0] + genD_T3[0] - conD_T3[0] #nmol/day

#TSH
n2_TSH = np.zeros(len(t))
n3_TSH = np.zeros(len(t))
genC_TSH = np.zeros(len(t))
n4_TSH = np.zeros(len(t))
conD_TSH = np.zeros(len(t))
n5_TSH = np.zeros(len(t))

n2_TSH[0] = SSTSH
n3_TSH[0] = n2_TSH[0]
genC_TSH[0] = 165.2
n4_TSH[0] = n3_TSH[0] + genC_TSH[0]
```

```python
conD_TSH[0] = 167.73
n5_TSH[0] = n4_TSH[0] - conD_TSH[0]
import pandas as pd
# Initialization Table
# Create a dictionary to store the initial flow rates
data = {
    "Constituent": [
        "Iodine (n1_I)", "Iodine (n2_I)", "Iodine (conB_I)", "Iodine (n3_I)", "Iodine (n4_I)",
        "Iodine (genD_I)", "Iodine (n6_I)", "Iodine (n5_I)", "Iodine (nstorage_I)",
        "T4 (n2_T4)", "T4 (genB_T4)", "T4 (n3_T4)", "T4 (n4_T4)", "T4 (conD_T4)", "T4 (n5_T4)",
        "T3 (n2_T3)", "T3 (genB_T3)", "T3 (n3_T3)", "T3 (n4_T3)", "T3 (genD_T3)", "T3 (conD_T3)", "T3 (n5_T3)",
        "TSH (n2_TSH)", "TSH (n3_TSH)", "TSH (genC_TSH)", "TSH (n4_TSH)", "TSH (conD_TSH)", "TSH (n5_TSH)"
    ],
    "Flow Rate (nmol/day)": [
        n1_I[0], n2_I[0], conB_I[0], n3_I[0], n4_I[0],
        genD_I[0], n6_I[0], n5_I[0], nstorage_I[0],
        n2_T4[0], genB_T4[0], n3_T4[0], n4_T4[0], conD_T4[0], n5_T4[0],
        n2_T3[0], genB_T3[0], n3_T3[0], n4_T3[0], genD_T3[0], conD_T3[0], n5_T3[0],
        n2_TSH[0], n3_TSH[0], genC_TSH[0], n4_TSH[0], conD_TSH[0], n5_TSH[0]
    ]
}

# Convert to a DataFrame
df = pd.DataFrame(data)

# Display the table
print(df.to_string(index=False))
```

```python
#print(I)
import numpy as np

MaxStorage = nstorage_I[0]
for i in range(1, len(t)):
    n_IN[i] = n_IN[i-1]

    # Streams 1 and 2
    n1_I[i] = n1_I[i-1]
    n2_T3[i] = n5_T3[i-1]
    n2_T4[i] = n5_T4[i-1]
    n2_TSH[i] = n5_TSH[i-1]
    n2_I[i] = n5_I[i-1] + n1_I[i-1]

    # Stream 3

    #Determination of Generation of T3 and T4 depending on TSH
    TheoreticalT4Gen = 119/237*n2_TSH[i]
    TheoreticalT3Gen = 7.68/237*n2_TSH[i]
    TheoreticalICon = 3 * TheoreticalT3Gen + 4 * TheoreticalT4Gen
    MaxICon = 0.2*n2_I[i]
    MaxT3Gen = MaxICon/65
    MaxT4Gen = MaxICon/4.2
    if (TheoreticalICon<MaxICon):
        genB_T4[i] = TheoreticalT4Gen
        genB_T3[i] = TheoreticalT3Gen
        conB_I[i] = TheoreticalICon

        if(nstorage_I[i-1]<MaxStorage):
            RefillTerm = 0.1*n2_I[i]
            if(nstorage_I[i-1] + RefillTerm> MaxStorage):
```

```python
            nstorage_I[i] = MaxStorage
            conB_I[i] += (MaxStorage-nstorage_I[i-1])

        else:
            nstorage_I[i] = nstorage_I[i-1] + RefillTerm
            conB_I[i] += RefillTerm
    else:
        nstorage_I[i] = nstorage_I[i-1]

else:
    if(nstorage_I[i-1]<=0): #Storage is empty
        genB_T4[i] = MaxT4Gen
        genB_T3[i] = MaxT3Gen
        #derive from the linear ratio between T3 and T4 generation
        #see appendix and math equations sources in reference
        conB_I[i] = MaxICon
        nstorage_I[i] = 0
    else:
        if(n2_TSH[i]>SSTSH):
            #TheoreticalISecretion = (n2_TSH[i]-237)/237*nstorage_I[i-1]
            TSH = n2_TSH[i]
            N = np.log(TSH)
            D = np.log(237)

            TheoreticalISecretion = N/D*100
            #the secretion follows a log relationship; we simplified the kinetics in existing
mathematical models to match our model
            #237 is avg TSH in blood flow to thyroid
        else:
            TheoreticalISecretion = 0
```

```python
    if(TheoreticalISecretion<nstorage_I[i-1]):
        genB_T4[i] = MaxT4Gen + TheoreticalISecretion/4.2
        genB_T3[i] = MaxT3Gen + TheoreticalISecretion/65
        conB_I[i] = MaxICon
        nstorage_I[i] = nstorage_I[i-1] - TheoreticalISecretion
        #print("Help:",TheoreticalISecretion)

    else:
        genB_T4[i] = MaxT4Gen + nstorage_I[i-1]/65
        genB_T3[i] = MaxT3Gen + nstorage_I[i-1]/4.2
        conB_I[i] = MaxICon
        nstorage_I[i] = 0

n3_T3[i] = n2_T3[i] + genB_T3[i]
n3_T4[i] = n2_T4[i] + genB_T4[i]
n3_TSH[i] = n2_TSH[i]
n3_I[i] = n2_I[i] - conB_I[i]

# Stream 4
genC_TSH[i] = 880953 / (n3_T4[i])
#880953 is the ratio relationship calculated from steady state

n4_T3[i] = n3_T3[i]
n4_T4[i] = n3_T4[i]
n4_TSH[i] = n3_TSH[i] + genC_TSH[i]
n4_I[i] = n3_I[i]

# Stream 5
# if(n4_TSH[i]>500):
#   conD_TSH[i] = conD_TSH[0]
#   genD_T3[i] = genD_T3[0]
```

```python
#   conD_T3[i] = conD_T3[0]

#   conD_T4[i] = conD_T4[0]

#   genD_I[i] = genD_I[0]

#else:

conD_TSH[i] = conD_TSH[0]

genD_T3[i] = 0.0072013*n4_T4[i]

conD_T4[i] = genD_T3[i] + 0.01511407*n4_T4[i]

conD_T3[i] = 0.1393285*n4_T3[i]

genD_I[i] = 3*conD_T3[i] + +4*0.01511407*n4_T4[i] + genD_T3[i]

n5_TSH[i] = n4_TSH[i] - conD_TSH[i]

n5_T3[i] = n4_T3[i] + genD_T3[i] - conD_T3[i]

n5_T4[i] = n4_T4[i] - conD_T4[i]

n6_I[i] = 0.59*n4_I[i]

if(n4_TSH[i]<0): #holds min stream flow at zero (can't have negative concentration)

  n4_TSH[i] = 0

#0.59: excretion is proportion of what comes in (comparing ratio n4 to n6)

n5_I[i] = n4_I[i] - n6_I[i] + genD_I[i]

if(n5_T3[i]<0):

  n5_T3[i] = 0

if(n5_T4[i]<0):

  n5_T4[i] = 0

# Plot the main variables

plt.figure(figsize=(10, 5))

plt.plot(t, n_IN, 'k--',label= 'Normal Iodide')

plt.plot(t, n2_I, 'b-', label='Iodide')

plt.plot(t, n5_T3, 'r-', label='T3')
```

```python
plt.plot(t, n5_T4, 'g-', label='T4')
#plt.plot(t, nstorage_I, 'k--', label='Iodide Storage')
plt.xlabel('Time (days)', fontsize=15)
plt.ylabel('nmoles/day', fontsize=15)
plt.legend()
plt.title('Iodide, T3, and T4 Flowrate in Severely Diseased State', fontsize=16)
plt.rc('xtick', labelsize=12)
plt.rc('ytick', labelsize=12)
plt.grid()
plt.show()

# Plot TSH separately (convert to mU by dividing by 237 and multiplying by 1000)
plt.figure(figsize=(10, 5))
plt.plot(t, n5_TSH, 'y-', label='TSH (mU)')
plt.xlabel('Time (days)')
plt.ylabel('mU')
plt.legend()
plt.title('TSH Levels')
plt.grid()
plt.show()

# Plot Storage separately
plt.figure(figsize=(10, 5))
plt.plot(t, nstorage_I, 'k--', label='Iodide Storage')
plt.xlabel('Time (days)')
plt.ylabel('nmol')
plt.legend()
plt.title('Iodide Storage')
plt.grid()
plt.show()
```

**References**

[1] A. Hatch-McChesney and H. R. Lieberman, "Iodine and Iodine Deficiency: A Comprehensive Review of a Re-Emerging Issue," *Nutrients*, vol. 14, no. 17, p. 3474, Aug. 2022. [Online]. Available: https://pmc.ncbi.nlm.nih.gov/articles/PMC9459956/.

[2] A. C. Schroeder and M. L. Privalsky, "Thyroid hormones, T3 and T4, in the brain," *Frontiers in Endocrinology*, vol. 5, p. 40, 2014. doi: 10.3389/fendo.2014.00040.

[3] "Understanding iodine deficiency," *Lakecountyin.gov*, 2024. [Online]. Available: https://lakecountyin.gov/departments/health/nursing-clinic/diseases-and-conditions/nutritional-conditions/understanding-iodine-deficiency. Accessed: Feb. 26, 2025.

[4] *The Nutrition Source*, "Iodine," *Harvard T.H. Chan School of Public Health*, Oct. 19, 2021. [Online]. Available: https://nutritionsource.hsph.harvard.edu/iodine/.

[5] L. Chaker, A. C. Bianco, J. Jonklaas, and R. P. Peeters, "Hypothyroidism," *The Lancet*, vol. 390, no. 10101, pp. 1550–1562, Sep. 2020, doi: 10.1016/s0140-6736(17)30703-1.

[6] M. Armstrong, E. Asuka, and A. Fingeret, "Physiology, Thyroid Function," *NIH.gov*, Jun. 28, 2019. [Online]. Available: https://www.ncbi.nlm.nih.gov/books/NBK537039/.

[7] World Health Organization, "Iodine deficiency," *WHO*, 2013. [Online]. Available: https://www.who.int/data/nutrition/nlis/info/iodine-deficiency.

[8] *PubChem*, "Iodide ion," *National Center for Biotechnology Information*. [Online]. Available: https://pubchem.ncbi.nlm.nih.gov/compound/Iodide-ion.

[9] A. Milanesi and G. A. Brent, "Iodine and Thyroid Hormone Synthesis, Metabolism, and Action," *ScienceDirect*, Jan. 01, 2017. [Online]. Available: https://www.sciencedirect.com/science/article/abs/pii/B9780128021682000129.

[10] M. A. Shahid, M. A. Ashraf, and S. Sharma, "Physiology, Thyroid Hormone," *StatPearls*, Jan. 2025. Available: https://pubmed.ncbi.nlm.nih.gov/29763182/ .

[11] "Peripheral Thyroid Hormone Conversion and Its Impact on TSH and Metabolic Activity - Restorative Medicine," Restorative Medicine, Apr. 2014. https://restorativemedicine.org/journal/peripheral-thyroid-hormone-conversion-and-its-impact-on-tsh-and-metabolic-activity/.

[12] R. Mullur, Y.-Y. Liu, and G. A. Brent, "Thyroid Hormone Regulation of Metabolism," Physiological Reviews, vol. 94, no. 2, pp. 355–382, Apr. 2014, doi: 10.1152/physrev.00030.2013.

[13] J. Jonklaas, "Optimal Thyroid Hormone Replacement," *Endocrine Reviews*, vol. 43, no. 2, Sep. 2021, doi: 10.1210/endrev/bnab031.

[14] H. R. Chung, "Iodine and thyroid function," Annals of Pediatric Endocrinology & Metabolism, vol. 19, no. 1, pp. 8–12, Mar. 2014. [Online]. Available: https://e-apem.org/upload/pdf/apem-19-8.pdf.

[15] M. E. Krass, F. S. Labella, and R. Mailhot, "Bovine Thyrotropin," *Endocrinology*, vol. 84, no. 5, pp. 1257–1261, May 1969, doi: 10.1210/endo-84-5-1257.

[16] L. Tegler et al., "Thyroid Blood Flow Rate in Man," J. Endocrinol. Investig., vol. 4, no. 3, pp. 335–341, 1981, doi: 10.1007/BF03349454.

[17] I. J. Chopra, "Assessment of Daily Production of Reverse T3 in Man," *J. Clin. Investig.*, vol. 58, no. 1, pp. 32–40, Jul. 1976, doi: 10.1172/jci108456.

[18] G. Lisco et al., "Interference on Iodine Uptake," *Nutrients*, vol. 12, no. 6, p. 1669, Jun. 2020, doi: 10.3390/nu12061669.

[19] "T3," *Tocris Bioscience*, 2017. [Online]. Available: https://www.tocris.com/products/t3_6666. Accessed: Feb. 25, 2025.

[20] "Thyroxine (T4 hormone human)," *Molecular Depot*, Jan. 19, 2025. [Online]. Available: https://moleculardepot.com/product/l-thyroxine-t4/. Accessed: Feb. 25, 2025.

[21] J. E. Hall and M. E. Hall, *Guyton and Hall Textbook of Medical Physiology*, 14th ed. Elsevier, 2021, p. 941.

[22] T. M. Wolff, C. Veil, J. W. Dietrich, and M. A. Müller, "Mathematical modeling and simulation of thyroid homeostasis: Implications for the Allan-Herndon-Dudley syndrome," *Frontiers in Endocrinology*, vol. 13, Dec. 2022. doi: 10.3389/fendo.2022.882788.

[23] R. Li, *Mathematical Modeling of Thyroid Hormone Metabolism in the Human Liver*, M.S. thesis, Emory Univ., Atlanta, GA, USA, 2021. [Online]. Available: https://etd.library.emory.edu/concern/etds/k930bz397?locale=es. Accessed: Feb. 24, 2025.

[24] L. H. Fish, H. L. Schwartz, J. Cavanaugh, M. W. Steffes, J. P. Bantle, and J. H. Oppenheimer, "Replacement dose, metabolism, and bioavailability of levothyroxine in the treatment of hypothyroidism," *New England Journal of Medicine*, vol. 316, no. 13, pp. 764–770, Mar. 1987. doi: 10.1056/nejm198703263161302.

[25] C. A. Spencer *et al.*, "Applications of a new chemiluminometric thyrotropin assay to subnormal measurement," *The Journal of Clinical Endocrinology & Metabolism*, vol. 70, no. 2, pp. 453–460, Feb. 1990. doi: 10.1210/jcem-70-2-453.

[26] W. D. Odell, R. D. Utiger, J. F. Wilber, and P. G. Condliffe, "Estimation of the secretion rate of thyrotropin in man," *The Journal of Clinical Investigation*, vol. 46, no. 6, pp. 953–959, Jun. 1967. doi: 10.1172/JCI105601.

[27] S. M. Abdalla and A. C. Bianco, "Defending plasma T3 is a biological priority," *Clinical Endocrinology*, vol. 81, no. 5, pp. 633–641, Aug. 2014. doi: 10.1111/cen.12538.

[28] N. B. Myant and E. E. Pochin, "The thyroid clearance rate of plasma iodine as a measure of thyroid activity," *Proceedings of the Royal Society of Medicine*, vol. 42, no. 12, pp. 959–961, Dec. 1949. doi: 10.1177/003591574904201202.

[29] National Institutes of Health, "Iodine - Health Professional Fact Sheet." [Online]. Available: https://ods.od.nih.gov/factsheets/Iodine-HealthProfessional/. Accessed: Mar. 03, 2025.